\documentclass[10pt,journal]{IEEEtranTCOM}

\usepackage[english]{babel}
\usepackage[usenames]{color}
\usepackage[cp1250]{inputenc}
\usepackage{amsfonts}
\usepackage{amsthm}
\usepackage{graphicx}
\usepackage{epsfig}
\usepackage{mathrsfs}
\usepackage{amsmath}
\usepackage{algorithm}
\usepackage{algorithmic}

\theoremstyle{plain}

\begin{document}
\newcommand{\bea}{\begin{eqnarray}}
\newcommand{\eea}{\end{eqnarray}}
\newcommand{\be}{\begin{equation}}
\newcommand{\ee}{\end{equation}}
\newcommand{\beas}{\begin{eqnarray*}}
\newcommand{\eeas}{\end{eqnarray*}}
\newcommand{\bs}{\backslash}
\newcommand{\bc}{\begin{center}}
\newcommand{\ec}{\end{center}}
\def\SC {\mathscr{C}}

\title{Distortion-Resistant Hashing \\for rapid search of similar DNA subsequence}
\author{\IEEEauthorblockN{Jarek Duda}\\
\IEEEauthorblockA{Jagiellonian University,
Golebia 24, 31-007 Krakow, Poland,
Email: \emph{dudajar@gmail.com}}}
\maketitle

\begin{abstract}
One of the basic tasks in bioinformatics is localizing a short subsequence $S$, read while sequencing, in a long reference sequence $R$, like the human geneome. A natural rapid approach would be finding a hash value for $S$ and compare it with a prepared database of hash values for each of length $|S|$ subsequences of $R$. The problem with such approach is that it would only spot a perfect match, while in reality there are lots of small changes: substitutions, deletions and insertions.

This issue could be repaired if having a hash function designed to tolerate some small distortion accordingly to an alignment metric (like Needleman-Wunch): designed to make that two similar sequences should most likely give the same hash value. This paper discusses construction of Distortion-Resistant Hashing (DRH) to generate such fingerprints for rapid search of similar subsequences. The proposed approach is based on the rate distortion theory: in a nearly uniform subset of length $|S|$ sequences, the hash value represents the closest sequence to $S$. This gives some control of the distance of collisions: sequences having the same hash value.
\end{abstract}
\section{Introduction}
DNA sequencing techniques can directly sequence a relatively short sequences: of 50-100000 nucleotides, referred as reads. In most of applications, these fragments need to be aligned in a known reference sequence, like a human genome with approximately 3 billions nucleotides. While there are very efficient methods for searching of subsequences, for example based on Burrows-Wheeler transform~\cite{BWT}, they require a perfect match for the subsequence we are looking for.

However, the real problem is that there are differences between reads and corresponding fragments of the reference sequence: substitution, insertions, deletions, or even translocation of large fragments. One reason are individual differences, which are usually the main purpose of sequencing: to provide personalized medicine for a given patient. The second reason are errors resultant from the sequencing technique, which are especially high for recent methods operating on single DNA strands, like PacBio or very promising Oxford nanopore sequencing~\cite{nanopore}. Hence, the practical methods are usually based on seek-an-extend approach: search for short subsequences of reads and then try to extend them using some dynamic programming method. This approach strongly depends on the way we choose the short subsequences of reads, which still might be too long to provide a perfect match, especially for high error rate methods like nanopore sequencing.

\begin{figure}[t!]
    \centering
        \includegraphics[width=8cm]{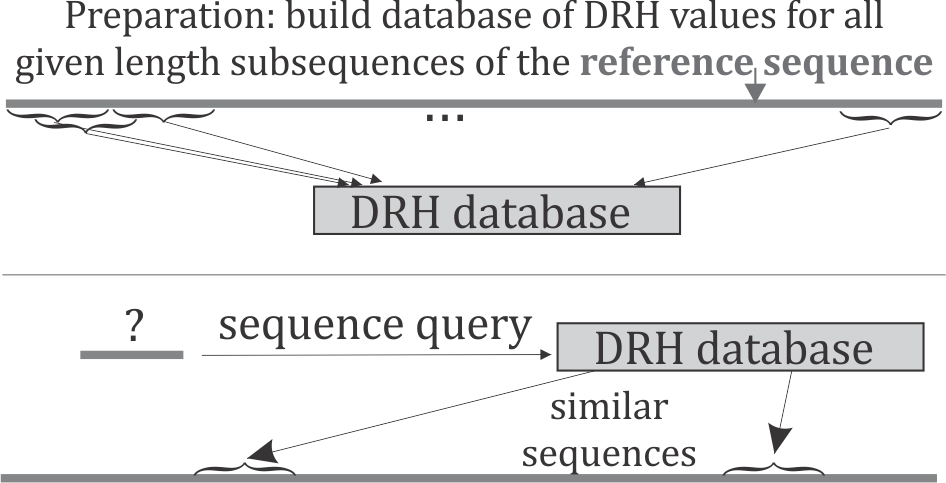}\
\begin{center}
        \caption{Scenario for rapid search of similar subsequences using DRH: first build database of DRH values for all given length subsequences of the reference sequence, then query returns positions of collisions: subsequences of the same DRH value. Using hash function instead we would look for a perfect match. In contrast, DRH tolerates some distortion of the sequence, accordingly to a chosen alignment metric and rate (the smaller rate, the larger tolerance).}
   \label{sash}
\end{center}
\end{figure}

This paper introduces Distortion-Resistant Hashing (DRH) to help with this difficulty. Like a hashing function, it returns a pseudoranandom value identifying a given object. However, in contrast to standard hash function, it has a special property that similar objects should get the same identifier, where similarity is defined by some metric, like Needleman-Wunch metric evaluating alignment of two sequences.

One of applications is constructing a database of such DRH values, for example using B-tree to get logarithmic search time for a large storage medium like HDD/SDD. They might correspond to all subsequences of some range of lengths of the reference sequence, like in Fig. \ref{sash}. Then we can find DRH value for a given read sequence and ask the database for positions of collisions: promising positions on the reference sequence.

\begin{figure}[t!]
    \centering
        \includegraphics[width=8cm]{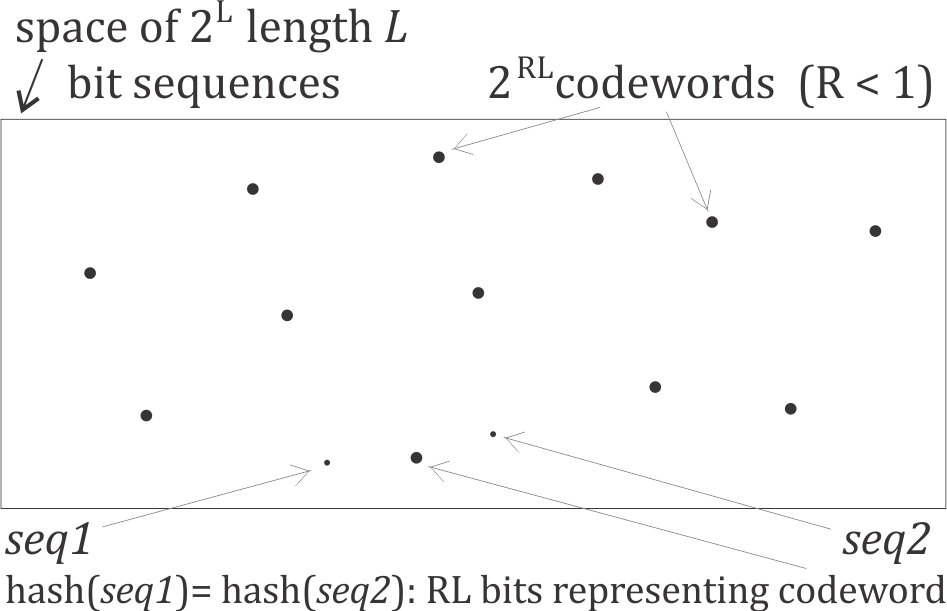}\
\begin{center}
        \caption{Rate distortion-based hashing: emphasize some (usually random) subset of all sequences: codewords, then hash value of a sequence is a representation of the closest codeword to this sequence. Assuming a metric was used for a distance and there is a bound for the distance from a codeword, triangle inequality allows to bound distance of two sequences corresponding to the same codewords (hash value).}
   \label{rate}
\end{center}
\end{figure}

The discussed approach is based on the rate distortion theory, which represents an object (sequence) as its lossy compression version: having a shorter representation, but being distorted. Figure \ref{rate} illustrates how it is realized: all sequences we can represent form a subset of all sequences. This subset is called a codebook and its instances are codewords. For example we can choose a $2^{RL}$ size pseudorandom subset of length $2^L$ bit sequences, for some rate $0<R<1$. This subset should cover the space in nearly uniform way, what is usually asymptotically guaranteed for choosing it in a pseudorandom way. Now we can encode a sequence as a representation ($LR$ bits) of the closest codeword. Using such representation as a hash value, similar sequences should have the same closest codeword, leading to the same hash value.

The length of such representation (DRH) depends on the length of the sequence - we can use separate databases for different lengths of DRH sequences. Alternatively, we can finally apply a standard hash function to these DRH sequences, getting a fixed length DRH values (for example 64 bit), what allows to merge these databases. There is an issue of having multiple close codewords for a given sequence, leading to lack of collision for similar sequences (false negatives) - it suggests using a few DRH values for each sequence: up to some distance difference.

The presented approach is very general, and have potential in many situations for searching for similar object, fragments, for example in analysis of text.

\section{Distortion-Resistant Hashing}
We will now discuss the concepts, practical construction and application of DRH. We will focus on working with nucleotide sequence $(\mathcal{A}=\{A,C,T,G\})$ and hash values being bit sequences. However, the considerations are very general, one could for example use alphabet for text.
\subsection{Needleman-Wunch alignment metric}
The standard way of evaluating alignment of two sequences use dynamic programming. We will focus on the simplest: Needleman-Wunch (NW) algorithm~\cite{needleman}, but this approach can be generalized for more complex ones. Its standard formulation searches alignment which maximizes a score. For better control, let us reformulate it into a metric: a function returning distance between two finite sequences $\left(d: \mathcal{A}^* \times \mathcal{A}^*\to \mathbb{R}\right)$ which for any $S,T,U\in\mathcal{A}^*$ fulfills:
\begin{itemize}
\item non-negativity: $d(S,T)\geq 0$,
  \item identity of indiscernibles: $d(S,T)=0 \Leftrightarrow S=T$,
  \item symmetry: $d(S,T)=d(T,S)$,
  \item trinagle inequality:   $d(S,T)+d(T,U)\geq d(S,U)$.
\end{itemize}

Now for a given two sequences $S,T\in\mathcal{A}^*$, define $M^{ST}_{ij}$ as alignment metric between length $i$ prefix of $S$ and length $j$ prefix of $T$, getting initial values and recurrence for dynamical programming:
$$M^{ST}_{0i}=M^{ST}_{i0}=i\cdot c_g\quad \textrm{for any }i\in\mathbb{N}$$
\be \textrm{for }i,j>0,\qquad M^{ST}_{ij}=  \label{rec}\ee
$$=\min(M^{ST}_{i,j-1}+c_g,M^{ST}_{i-1,j}+c_g,M^{ST}_{i-1,j-1}+c_s [S_i\neq T_i])$$
where $c_g>0$ is the cost of a gap, $c_s>0$ is cost of substitution and $[cond]=1$ if $cond$ is true, 0 otherwise.

Finally we get a natural definition:
\be d(S,T):=M^{ST}_{|S|,|T|} \ee
which, as it is easy to check, fulfills conditions for being a metric (triangle inequality comes from (\ref{rec})).

Optimal $c_g,c_s$ parameters can be chosen as logarithm of probabilities from statistical models for errors and for the sequence. It might be beneficial to make $c_s$ dependent on substitution type. One might also consider a more sophisticated gap scoring, as NW assumes exponential drop of probability of gap length. There is commonly used Waterman-Smith (WS) variant~\cite{waterman}, which is more tolerant for long gaps (affine-gap scoring), and can be analogously reformulated as metric. There can be also considered word-based distortions, like substituting a word with a synonym, or short DNA fragment with reverse-complementary sequence.

Using a metric is not necessary, but allows for a better control. For practical reason (search for the DRH), it is crucial that we can cheaply elongate the considered sequences - alignment scores requiring two-directional travelling over sequences would be much more expensive.

\subsection{Hashing based on rate distortion}
The situation of using rate distortion approach for hashing is schematically presented in Fig. \ref{rate}: in the space of $|\mathcal{A}|^L$ length $L$ sequences, we emphasize $|\mathcal{A}|^{RL}$ codewords. Hash value for a sequence is a unique representation of its closest codeword, what requires $$RL \log_2 (|\mathcal{A}|)\quad\textrm{bits of information (length of hash value)}$$
We will discuss choosing codewords in a pseudorandom way, what asymptotically leads to optimal capacity, as in the original discussion of Shannon. However, as short sequences are also often of interest, a dedicated codebook might be required in this case.

\begin{figure}[t!]
    \centering
        \includegraphics[width=6.5cm]{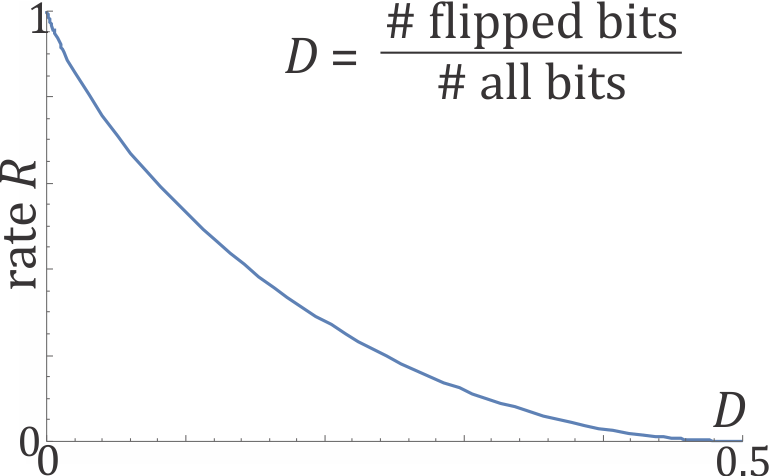}\
\begin{center}
        \caption{The distance between rate $R$ and distortion $D$ for binary alphabet and normalized Hamming distance.}
   \label{rbin}
\end{center}
\end{figure}

For standard metrics and asymptotic case, nearly all sequences (so called typical) have the same distance to the closest codeword: $D(R)$, which depends on the rate $R$. For example imagine simple substitution-only binary alphabet case with normalized Hamming distance: the number of changed bits divided by length. The number of length $L$ sequences in distance $D$ from a given sequence is ${L \choose DL}\approx 2^{L h(D)}$, where $h(p)=-p \log_2(p)- (1-p) \log_2 (1-p)$ is the Shannon entropy. Intersections of such balls in sequence space turn out asymptotically negligible, hence the number of codewords is asymptotically $2^L/2^{Lh(D)}=2^{L(1-h(D))}$, getting $R(D)=1-h(D)$ relation between rate and distance $D=D(R)$, presented in Fig. \ref{rbin}.

For finite sequences the situation is more complex and depends on the choice of codebook, but generally the distance from the nearest codeword forms a probability distribution around the $D(R)$ value, approaching Dirac delta in $D(R)$ for $L\to\infty$.

Allowing gaps: insertion-deletion errors, brings us to the realm of synchronization channels, where not even asymptotic situation is known in simplest cases like deletion channel. \\

Assuming there is some bound $D_{max}$ for distance between a sequence and its codeword, the triangle inequality would allow us to bound distance between two sequences, especially when they collide: have the same hash value.
\begin{itemize}
  \item if $hash(S)=hash(T)$ then $d(S,T) \leq 2 D_{max}$,
  \item generally $d(S,T) \leq 2D_{max} + d(S',T')$ where $S',\ T'$ are codewords corresponding to $S,\ T$.
\end{itemize}
As discussed, for finite sequences we should rather expect a probability distribution for $D$, allowing to write above inequalities with some certainty. This probability distribution can be approximated by simulations: generate many random sequences, test their distance to the corresponding codeword, and create a histogram.
\subsection{Search tree for the DRH value}
\begin{figure}[t!]
    \centering
        \includegraphics[width=8cm]{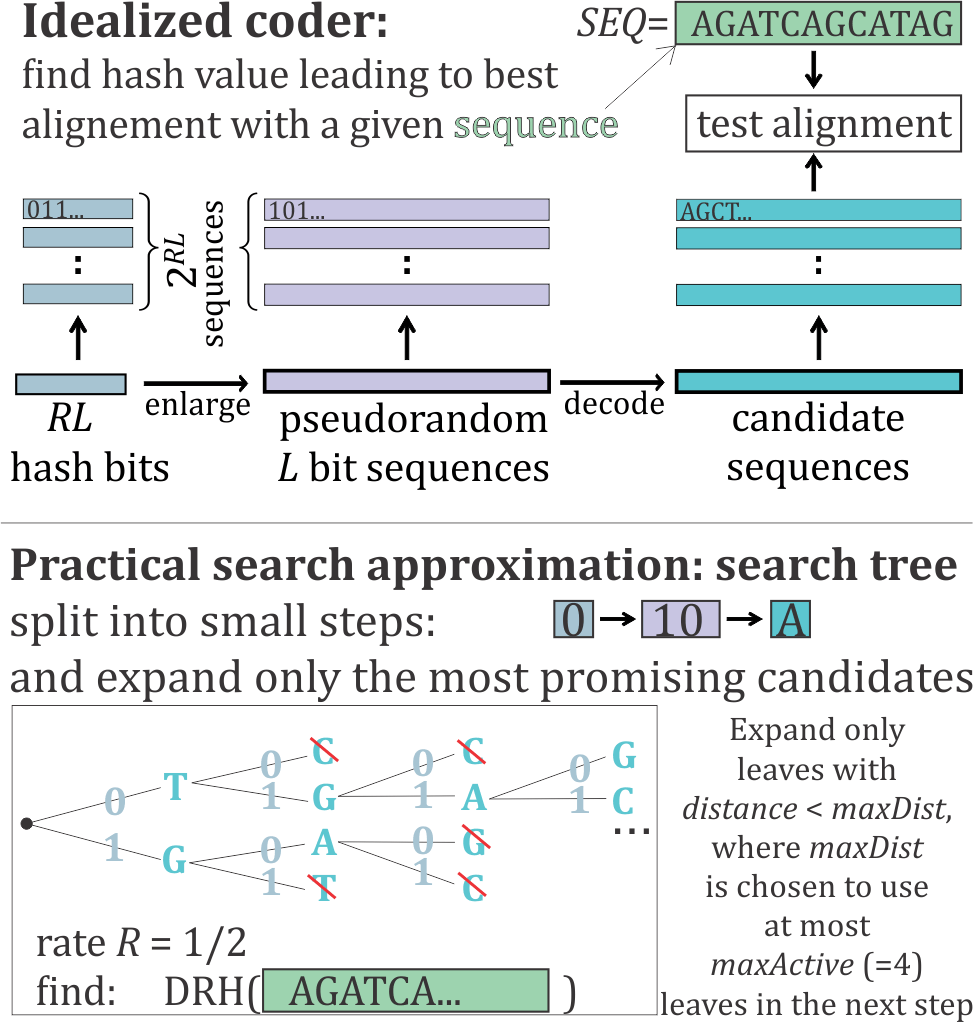}\
\begin{center}
        \caption{Top: idealized scheme for finding the DRH sequence: leading to a candidate having the best alignment with a given sequence $SEQ$. Bottom: practical realization by dividing into short blocks (a few bits in practice) and expanding only some number of the most promising leaves. In a given step we are focusing on nodes having the same $depth$. The symbols considered in a given moment should depend on the entire path, not only the last branch.}
   \label{schem}
\end{center}
\end{figure}
As depicted in Fig. \ref{schem}, an idealized way for finding the closest codeword would be generating all of them and calculating the distance. However, it is completely impractical in most of cases. For simple metrics there are possible LDPC-based methods to rate distortion~\cite{LDPC}, however, alignment metrics seem too complex for this kind of approaches.

We will discuss search tree approximation which was successfully applied for a related generalization of the Kuznetsov-Tsybakov problem~\cite{KT}, and can be adapted for complex alignment metrics. It is analogous to sequential decoding for error correction, allowing to efficiently work with synchronization problem like here, for example deletion channel~\cite{del}. Its modification for the DRH problem is schematically presented as Algorithm \ref{search} and is explained in the following paragraphs. Its $en\_dec$ is combination of enlarge and combine functions from Fig. \ref{schem}: it transforms blocks of hash function into blocks of codewords - sequences which should be close to $SEQ$. The choice of this function will be discussed in the following subsection.

This approach builds a tree of prefixes of codewords, expanding only the most promising ones (with good alignment). We divide the sequence into short blocks and in step $depth$ we consider only $depth$ first blocks of $SEQ$ sequence. In this step we use the best found prefixes of $depth-1$ blocks and try to expand them by one block in all possible ways. This way the formed tree has maximal degree $2^{block\_size}$ where $block\_size$ is the number of bits in block corresponding to hash sequence.

As the tree would grow exponentially, we need to add a mechanism to expand only the most promising nodes: having the shortest distance to $SEQ$. A simple way to do it, is finding a $maxDist$ threshold for expanding a depth $depth-1$ nodes in $depth$ step, such that the number of nodes in this step does not exceed $maxActive$ parameter (for example 100). The $maxDist$ can be approximated in linear time using buckets, like in \cite{PRcodes}. Parameter $maxActive$ controls time and memory requirement of the search, approaching the idealized search for $maxActive\to\infty$, for example the theoretical capacity in the generalized Kuznetsov-Tsybakov case~\cite{KT}.

Step $depth$ extends sequences of $depth-1$ blocks by a single block, dynamic programming is required to find distances for such extensions. For Needleman-Wunch metric there was used recurrence formula for $M^{ST}_{ij}$ matrix (\ref{rec}), which stores distance between length $i$ prefix of $S$ sequence and length $j$ prefix of $T$ sequence. Nodes of the search tree need to store the entire row of such $M$ matrix: $dist$ is 1D table, $dist[i]$ is alignment distance between the considered sequence and length $i$ prefix of $SEQ$. The function $dyn(dist,seq)$ starts with $dist$ table (row of $M$) for $depth-1$ block prefix, and expands it using (\ref{rec}) recurrence by $seq$ sequence, finally getting $dist$ table for the current $depth$ block prefix. For evaluating a given node, we can use the best alignment with a prefix of $SEQ$: $min(dist)$. For the reason of memory saving, instead of holding the entire row of $M$ matrix, $dist$ can hold only some number of values around the minimal distance position.

As the active nodes are costly due to $dist$ table, and we need to be able to recreate path, Algorithm \ref{search} uses $history$ inexpensive nodes to be finally able to reconstruct paths in the tree. Such nodes contain $hash$ value which is a bit block added to hash sequence by this node. The $previous$ and $curr$ tables contain active nodes: $curr$ are built by expanding $previus$ by a single block. The number of these active nodes is limited by $maxActive$.

\begin{algorithm}[t]

\footnotesize{

\caption{DRH search tree for $SEQ$ sequence}
\label{search}
\begin{algorithmic}
\STATE // $previous$, $current$ are active nodes to expand in this step
\STATE // $history$ stores tree to finally reconstruct the best path(s)
\STATE // $decode(state,hash)$ finds new sequence block and state
\STATE // $dyn(dist,seq)$ dynamic programming aligning with $SEQ$
\STATE $currentHist=0$; $block\_val = 1 << block\_size - 1$; 
\STATE $previous\_nodes=1$; $init(previous[1])$;\qquad // root of tree
\FOR{$depth =1 \to number\_of\_blocks$}
\STATE Find $maxDist$ to use at most $maxActive$ nodes in this step
\STATE $nCurr = 0$
\FOR{$i=1 \to previous\_nodes$}
\IF{$previous[i].minDist < maxDist$}
\STATE $history[currentHist].parent=previous[i].parent$
\STATE $history[currentHist].hash=previous[i].hash$
\STATE $currentState=previous[i].state$
\STATE $currDist=previous[i].dist$\qquad//distance table
\FOR{$hash=0 \to block\_size$}
\STATE $nCurr++$; \qquad //create node to consider
\STATE $(currSeq, newState) = en\_dec(currState,hash)$
\STATE $curr[nCurr].parent=currHist$
\STATE $curr[nCurr].state=newState$
\STATE $curr[nCurr].dist=dyn(currDist,currSeq)$
\STATE $curr[nCurr].minDist=min(curr[nCurr].dist)$
\STATE $curr[nCurr].hash=hash$
\ENDFOR
\ENDIF
\STATE $currentHist++$
\ENDFOR
\STATE $previous=current;\ previous\_nodes=nCurr$
\ENDFOR
\STATE Find leaf with minimal $minDist$ (or leaves us to some distance)
\STATE travel $history$ from this leaf to root, concatenating $hash$ blocks
\end{algorithmic}
}
\end{algorithm}

\subsection{Transforming hash into sequences}
There has remained to discuss the $en\_dec$ function which combines enlarge and decode from Fig. \ref{schem}: transforms $hash$ block of hash sequence, into a block of sequence compared with $SEQ$, for example of nucleotides.

A simplest approach is directly defining a $hash\to seq$ function, for example in a pseudorandom way. However, it would restrict to codewords made of some characteristic blocks, what could reduce performance.

To use more random codewords, we can use a $state$ dependent on the history, which determines new sequence blocks. We can adapt the  way from \cite{KT} for this purpose. Specifically, choose a pseudorandom $t: [0,block\_val]\to [0,2^{16}-1]$ function (assuming 16 bit $state$). The transition function is
$$newState= (state \oplus t(hash)) << n$$
where $<<n$ is cyclic bit-shift left by $n$, being the block size after enlarging, and $n$ youngest bits of this $newState$ is enlarged $hash$. For example $n=8$ and $block\_size \in\{4,5,6,7\}$ for rates $R=block\_size/n\in\{1/2,5/8,3/4,7/8\}$. Then decoding can be make by just taking pairs of bits of this $n$ bit block, and translating them into $\{A,C,G,T\}$ alphabet, getting 4 nucleotides per block for $n=8$.\\

The above approach produces final sequence with i.i.d. probability distribution with equal probabilities of all elements of alphabet. If the sequences we are operating on are distant from such simple probability distribution, we can use entropy coder to operate on codewords from the assumed statistical model. Huffman coding assumes probabilities being power of 1/2, what is a strong limitation. Range coding allows to operate on large alphabet using nearly accurate probabilities, but is relatively costly, requires multiplication. Recent tANS entropy coding~\cite{ANS} also allows to operate on alphabet using nearly accurate probabilities, however its decoding step is just ($X$ is state):
$$t=decodingTable[X]; symbol=t.symbol; $$
$$ X=t.newX + readBits(t.nbBits);$$
which is perfect for DRH application. For this purpose, we need to prepare $decodingTable$ for probability distribution on alphabet of for example 4 successive nucleotides (256 possibilities). Then enlarged $hash$ for a given node are $nbBits$ bit sequences - for example $hash$ are all $nbBits-1$ bit sequences, and enlarging just inserts zero on some position.
\subsection{Practical remarks}
The DRH sequence produced by the discussed algorithm has length proportional to the length of the original sequence (is $\approx R$ times shorter), while in practice we often want to work on constant length hash values. This issue can be handled by just using some standard hash function on the DRH sequence, getting a fixed length DRH value having nearly the same properties. Using various length DRH sequences has advantage for initial selection of length: we can use separate smaller databases corresponding to a given DRH length.

The rate distortion approach divides the space of sequences into kind of Voronoi regions corresponding to each codeword. We would like that two close sequences get the same DRH, however, it might be not true when they are close to the boundary between some two regions. A solution for this false negativity problem is generating multiple codewords (hash values) for a given sequence: for example having a distance smaller than the minimal distance plus some small parameter. In the case of Algorithm \ref{search}, we finally get at most $maxActive$ candidates: instead of considering only the best of them, we can take some number of the lowest distance codewords, up to some difference from the optimal found. Using multiple DRH for a sequence, there is collision between two sequences (similarity) when they share at least one DRH. Hence, while building a database, all such hash values are inserted. While querying, we ask the database for each of generated DRH.

There have remained complex questions of optimal choice of parameters, like lengths of fragments to ask for DRH collisions. Thanks to distortion-resistance, they can be longer than for a perfect match. However, using too long might make us miss chimeric reads or split alignment. Considering a range of lengths seems reasonable. Another very important question is choosing the rate: larger rate means smaller tolerance for distortion.

The discussed search was for relatively simple Needleman-Wunch distance. Its expansion for affine gap scoring would require additional information in the $dist$ table about the last gap. There could be also used more complicated distortions like replacing a word with a synonym, or DNA fragment with reverse-complementary sequence. While the discussed algorithm for performance reasons was testing only the newly added block, one could test the entire generated prefix, allowing to consider much more complex distortion families.

\section{Conclusions and other applications}
There was presented and discussed a general concept of DRH: rate distortion-based hashing function: which doesn't change for small distortion of the object. It allows for rapid search for similar sequences by looking for collisions in database of DRH values. It seems a valuable tool for various situations when we need to search for similar sequences, for example:
\begin{itemize}
  \item various sequence alignment problems, especially in bioinformatics, 
  \item searching for similar fragments of text for various text analysis tasks, like web search or plagiarism checking,
  \item fractal compression of text: finding and pointing a similar fragment of text, then encode only the difference.     
\end{itemize}
\bibliographystyle{IEEEtran}
\bibliography{cites}

\begin{thebibliography}{1}
\providecommand{\url}[1]{#1}
\csname url@samestyle\endcsname
\providecommand{\newblock}{\relax}
\providecommand{\bibinfo}[2]{#2}
\providecommand{\BIBentrySTDinterwordspacing}{\spaceskip=0pt\relax}
\providecommand{\BIBentryALTinterwordstretchfactor}{4}
\providecommand{\BIBentryALTinterwordspacing}{\spaceskip=\fontdimen2\font plus
\BIBentryALTinterwordstretchfactor\fontdimen3\font minus
  \fontdimen4\font\relax}
\providecommand{\BIBforeignlanguage}[2]{{%
\expandafter\ifx\csname l@#1\endcsname\relax
\typeout{** WARNING: IEEEtran.bst: No hyphenation pattern has been}%
\typeout{** loaded for the language `#1'. Using the pattern for}%
\typeout{** the default language instead.}%
\else
\language=\csname l@#1\endcsname
\fi
#2}}
\providecommand{\BIBdecl}{\relax}
\BIBdecl

\bibitem{BWT}
H.~Li and R.~Durbin, ``Fast and accurate short read alignment with
  burrows--wheeler transform,'' \emph{Bioinformatics}, vol.~25, no.~14, pp.
  1754--1760, 2009.

\bibitem{nanopore}
M.~Jain, I.~T. Fiddes, K.~H. Miga, H.~E. Olsen, B.~Paten, and M.~Akeson,
  ``Improved data analysis for the minion nanopore sequencer,'' \emph{Nature
  methods}, vol.~12, no.~4, pp. 351--356, 2015.

\bibitem{needleman}
S.~B. Needleman and C.~D. Wunsch, ``A general method applicable to the search
  for similarities in the amino acid sequence of two proteins,'' \emph{Journal
  of molecular biology}, vol.~48, no.~3, pp. 443--453, 1970.

\bibitem{waterman}
T.~F. Smith and M.~S. Waterman, ``Identification of common molecular
  subsequences,'' \emph{Journal of molecular biology}, vol. 147, no.~1, pp.
  195--197, 1981.

\bibitem{LDPC}
E.~Martinian and M.~Wainwright, ``Low density codes achieve the rate-distortion
  bound,'' in \emph{Data Compression Conference, 2006. DCC 2006.
  Proceedings}.\hskip 1em plus 0.5em minus 0.4em\relax IEEE, 2006, pp.
  153--162.

\bibitem{KT}
J.~Duda, P.~Korus, N.~Gadgil, K.~Tahboub, and E.~J. Delp, ``Image-like 2d
  barcodes using generalizations of the kuznetsov-tsybakov problem.''

\bibitem{del}
J.~Duda, https://github.com/JarekDuda/DeletionChannelPracticalCorrection.

\bibitem{PRcodes}
------, https://github.com/JarekDuda/PRcodes.

\bibitem{ANS}
J.~Duda, K.~Tahboub, N.~J. Gadgil, and E.~J. Delp, ``The use of asymmetric
  numeral systems as an accurate replacement for huffman coding,'' in
  \emph{Picture Coding Symposium (PCS), 2015}.\hskip 1em plus 0.5em minus
  0.4em\relax IEEE, 2015, pp. 65--69.

\end{thebibliography}
\end{document}